\documentclass{article}
\usepackage{spconf,amsmath}
\usepackage[dvipdfmx]{graphicx}

\usepackage{dcolumn}
\usepackage{lipsum}
\usepackage{xurl}
\usepackage{color}
\usepackage[nobreak]{cite}
\usepackage{siunitx}
\graphicspath{{./figures/}}
\newcommand\figref[1]{Figure~\ref{#1}}
\newcommand\tabref[1]{Table~\ref{#1}}
\newcommand\secref[1]{Section~\ref{#1}}

\usepackage{array}
\newcommand{\PreserveBackslash}[1]{\let\temp=\\#1\let\\=\temp}
\newcolumntype{C}[1]{>{\PreserveBackslash\centering}p{#1}}
\newcolumntype{R}[1]{>{\PreserveBackslash\raggedleft}p{#1}}
\newcolumntype{L}[1]{>{\PreserveBackslash\raggedright}p{#1}}

\usepackage{caption}
\makeatletter
\newcommand\aepath[1]{
  \bgroup
    \ttfamily
    \ae@path#1\relax\@nil
  \egroup}
\def\ae@path#1#2\@nil{
  \def\ae@continue{}
  \detokenize{#1}\unskip\penalty\z@
  \ifx\relax#2
  \else
    \def\ae@continue{\ae@path#2\@nil}
  \fi
  \ae@continue}
\makeatother
\let\texttt\aepath

\title{Construction of a Large-scale Japanese ASR Corpus on TV Recordings}
\name{
  Shintaro Ando$^{1,2}$,
  Hiromasa Fujihara$^{1}$
  \thanks{We thank NTT PC Communications Inc. and their Innovation LAB program for letting us use their GPU resources.}
}
\address{$^{1}\,$Laboro.AI, Inc., Tokyo, Japan \\
$^{2}\,$Graduate School of Engineering, The University of Tokyo, Tokyo, Japan \\
\{ando,fujihara\}@laboro.ai
}

\begin{document}
\ninept
\baselineskip 0.97\normalbaselineskip
\maketitle
\begin{abstract}
This paper presents a new large-scale Japanese speech corpus for training automatic speech recognition (ASR) systems.
This corpus contains over 2,000 hours of speech with transcripts built on Japanese TV recordings and their subtitles.
We develop herein an iterative workflow to extract matching audio and subtitle segments from TV recordings based on a conventional method for lightly-supervised audio-to-text alignment.
We evaluate a model trained with our corpus using an evaluation dataset built on Japanese TEDx presentation videos and confirm that the performance is better than that trained with the Corpus of Spontaneous Japanese (CSJ).
The experiment results show the usefulness of our corpus for training ASR systems.
This corpus is made public for the research community along with Kaldi scripts for training the models reported in this paper.
\end{abstract}
\begin{keywords}
  Automatic speech recognition,
  Corpus
\end{keywords}
\section{Introduction}
\label{sec:intro}

Deep learning-based automatic speech recognition (ASR) systems heavily rely on the quantity of training data.
Many studies today, especially those for commercial purposes, use over 5000 hours of speech data \cite{jaitlyApplicationPretrainedDeep2012,amodeiDeepSpeechEndtoEnd2016,parthasarathiLessonsBuildingAcoustic2019}.
However, the amount of publicly available speech corpora is insufficient for languages other than English.
While it is common in English ASR research to use relatively large corpora such as the SwitchBoard-Fisher dataset and LibriSpeech\cite{panayotovLibrispeechASRCorpus2015}, each containing 2,000 and 960 hours of speech, respectively, the sizes of the commonly-used Japanese speech corpora are much smaller.
For example, two of the most popular corpora in Japanese ASR research, namely the Corpus of Spontaneous Japanese (CSJ) \cite{maekawaCorpusSpontaneousJapanese2003} and the Japanese Newspaper Article Sentences (JNAS) corpus \cite{itouJNASJapaneseSpeech1999}, only contain approximately 600 and 90 hours of speech, respectively.

Although an increasing demand for large-scale speech corpora for training high-quality ASR systems has been observed, constructing them is not straightforward because it usually requires massive manual labor of transcription or recording.
Many studies have been conducted to collect data without such laborious manual works.
One approach to solving the problem is to collect data in a community-driven manner, as done in the VoxForge\cite{VoxForge} and the Common Voice\cite{ardilaCommonVoiceMassivelyMultilingual2020} project.
Another common solution is to take advantage of the existing data and format them as a speech corpus.
For example, the aforementioned LibriSpeech \cite{panayotovLibrispeechASRCorpus2015} corpus was built from publicly available audiobooks.
Building an ASR corpus using social media videos with manual subtitles, which is less certain as transcripts, has also been investigated.
The TED-LIUM corpus\cite{hernandezTEDLIUMTwiceMuch2018}, containing 450 hours of speech, was built from TED conference videos.
Soltau {\it et al.} \cite{soltauNeuralSpeechRecognizer2017} extracted more than 100,000 hours of training data from YouTube videos, although they are not publicly available.

Broadcast TV programs with subtitles can be another source of speech corpora, and their usage has been investigated.
The timestamps of TV subtitles are often inaccurate; hence, several studies have been made to correct them.
The {\it Alignment} task of the Multi-Genre Broadcast (MGB) Challenges\cite{MGBChallenge, bellMGBChallengeEvaluating2015} required researchers to train acoustic models from TV recordings by providing them with English and Arabic TV recordings.
Several research teams explored the effect of forced alignment and lightly-supervised decoding\cite{bellMGBChallengeEvaluating2015, bellSystemForAutomaticAlignment2015, lanchantinSelectionOfMultiGenre2016, manoharJHUKaldiSystem2017}.
Bang {\it et al.} \cite{bangAutomaticConstructionLargeScale2020} built a Korean ASR corpus containing 336 hours of speech from a Korean TV broadcast, where the authors focused on utilizing the original timestamps and refining them effectively.

Following the studies that utilized broadcast TV programs, this paper introduces LaboroTVSpeech, a new large-scale Japanese ASR corpus built on Japanese TV recordings and their subtitles.
The current release contains approximately 2,000 hours of speech extracted from 8 months of recording period.
The process of corpus construction is fully automated; thus, the amount of data available will increase as time goes by.

The main contributions of this paper are as follows:
(1) We established an automated workflow for constructing an ASR corpus from a Japanese TV broadcast with inaccurate subtitles.
(2) We developed a new Japanese speech corpus, called LaboroTVSpeech, which contains 2,000 hours of speech, and made it publicly available for the research community\footnote{\url{https://github.com/laboroai/LaboroTVSpeech}}. To the best of our knowledge, there is no publicly-available Japanese corpus of this scale.
(3) We also established an evaluation dataset for ASR systems, called TEDxJP-10K, which is independent from the other ASR corpora and lets us fairly evaluate the performance of the ASR corpora.

The rest of the paper is organized as follows.
Section 2 elaborates on the corpus construction procedure from the original TV recordings and their subtitles.
Section 3 describes the statistical details of the corpus.
Section 4 presents the usefulness of the corpus through ASR experiments; and Section 5 concludes the paper.

\section{Corpus Construction Procedure}
\label{sec:collection}

\begin{figure*}
  \centering
  \includegraphics[width=\linewidth]{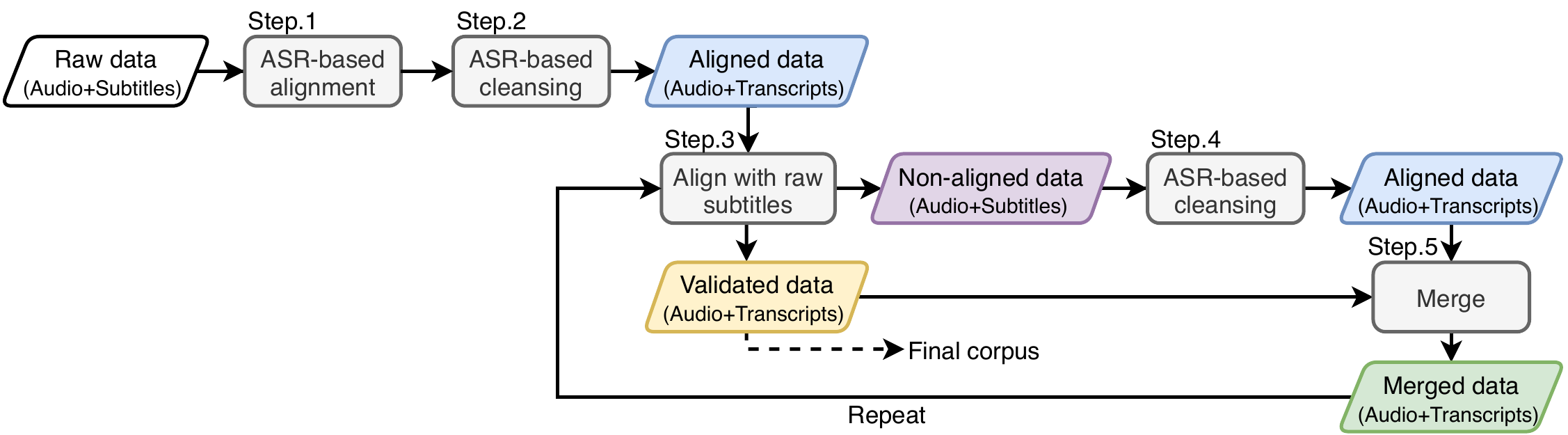}
  \caption{Diagram of our framework.}
  \label{fig:diagram}
  \vspace{-1.5\baselineskip}
\end{figure*}

\subsection{Data Collection and Preprocessing}
\label{ssec:collection}

The original audio data, their subtitles, and basic information such as the title and genre of each TV program, were collected from {\it 1seg} broadcast, which is a Japanese terrestrial digital audio/video and data broadcasting service for mobile devices.
{\it 1seg} broadcast signals are not protected by access control and legally accessible by any receivers, as opposed to ordinary broadcast TV signals that requires a special card named {\it B-CAS} containing the cryptography key to decode and are only accessible to authorized receivers.
{\it 1seg} broadcast contains basically same programs with ordinary broadcast signals but with a lower bit rate, although their difference is negligible when downsampled to \SI{16}{kHz}.
The current corpus includes recordings over the period of February--September 2020.
We ignored TV programs that are reruns of already-recorded ones and those without subtitles.

We recorded 9,142 TV programs of various lengths.
Some information on the details of the original recording is further reported in \secref{sec:corpus_details}.
Audio signals were decompressed and downsampled to \SI{16}{kHz}.
Each subtitle segment was tokenized into a sequence of words using MeCab\cite{kudoMeCabAnotherPartofSpeech} with mecab-ipadic-NEologd\cite{satoNeologismDictionaryBased2015} as a dictionary.
MeCab does not correctly tokenize numbers with more than a digit written in Arabic numerals; hence, we converted Arabic numerals to Chinese characters and re-inferred their pronunciations, if possible.
We also added several Japanese-style pronunciations of English words to our dictionary using the dictionary file released in the Bilingual Emacspeak Project \cite{BilingualEmacspeakProject}.

\begin{table*}
  \center
  \caption{
    Statistics of the raw recordings and the extracted corpus.
    The numbers shown in the parentheses in the audio lengths denote the percentages of each genre accounted for in the final extracted data.
  }
  \label{tab:genre_statistics}
  \begin{tabular}{l||rrR{0em}R{2em}L{1em}|rrc}
    \hline
                                    & \multicolumn{5}{c|}{Audio length (hours)} & \multicolumn{3}{c}{\#Words in trascripts}                                                                                                                                   \\
    \cline{2-6}
    Genre                           & \multicolumn{1}{c}{Raw}                   & \multicolumn{4}{c|}{Extracted}            & \multicolumn{1}{c}{Raw} & \multicolumn{1}{c}{Extracted} & \multicolumn{1}{c}{Extraction rate (\%)}                              \\
    \hline
    News, report                    & 1800.0                                    & 767.0                                     & (                       & 37.4                          & \%)                                      & 9,380 K  & 8,436 K  & 89.9 \\
    Variety show                    & 1382.9                                    & 316.4                                     & (                       & 15.4                          & \%)                                      & 6,330 K  & 3,368 K  & 53.2 \\
    Information/tabloid show        & 820.8                                     & 323.3                                     & (                       & 15.8                          & \%)                                      & 4,812 K  & 3,604 K  & 74.9 \\
    Drama                           & 664.0                                     & 206.0                                     & (                       & 10.1                          & \%)                                      & 3,175 K  & 2,304 K  & 72.6 \\
    Documentary/culture             & 462.4                                     & 175.6                                     & (                       & 8.6                           & \%)                                      & 2,158 K  & 1,780 K  & 82.5 \\
    Hobby/education                 & 304.1                                     & 101.0                                     & (                       & 4.9                           & \%)                                      & 1,613 K  & 1,092 K  & 67.7 \\
    Sports                          & 223.9                                     & 66.6                                      & (                       & 3.2                           & \%)                                      & 1,140 K  & 726 K    & 63.7 \\
    Animation/special effect movies & 175.2                                     & 39.7                                      & (                       & 1.9                           & \%)                                      & 801 K    & 430 K    & 53.7 \\
    Music                           & 99.2                                      & 17.6                                      & (                       & 0.9                           & \%)                                      & 281 K    & 184 K    & 65.6 \\
    Welfare                         & 58.7                                      & 20.1                                      & (                       & 1.0                           & \%)                                      & 286 K    & 212 K    & 74.0 \\
    Movies                          & 28.4                                      & 6.2                                       & (                       & 0.3                           & \%)                                      & 121 K    & 72 K     & 59.4 \\
    Theatre/public performance      & 23.5                                      & 10.4                                      & (                       & 0.5                           & \%)                                      & 150 K    & 118 K    & 78.5 \\
    \hline
    Total                           & 6042.9                                    & 2049.9                                    & (                       & 100.0                         & \%)                                      & 30,253 K & 22,331 K & 73.8 \\
    \hline
  \end{tabular}
  \vspace{-.5\baselineskip}
\end{table*}

\begin{table}
  \center
  \caption{
    Data subsets in LaboroTVSpeech.
  }
  \label{tab:corpus_splits}
  \begin{tabular}{l||rr}
    \hline
                         & \multicolumn{1}{c}{train} & \multicolumn{1}{c}{dev} \\
    \hline
    Audio length (hours) & 2036.2                    & 13.7                    \\
    \# Audio segments    & 1.6 M                     & 12 K                    \\
    \# Words             & 22 M                      & 147 K                   \\

    \hline
  \end{tabular}
  \vspace{-1.5\baselineskip}
\end{table}

\subsection{Audio and Subtitle Alignment}
\label{ssec:align}

TV programs and their subtitles have the following hallmarks, which make both subtitle texts and timestamps unreliable.
\begin{itemize}
  \item The subtitle timestamps are not always on time. For programs with real-time subtitles, such as news reports, they could have tens of seconds of delay. Even non-realtime subtitles could deviate from actual utterances for a few seconds.
  \item Even when programs themselves have subtitles, some audio segments (i.e., commercials) do not have them.
  \item Subtitles are not always verbatim; words may be formatted in a simpler form for readability. 
  In the Japanese language, spoken words and their formatted written version can be largely different. 
  \item In other cases, some parts of words in utterances are shown in the screen as images and excluded in subtitle texts. In such cases, if you listen to audio and read subtitles without watching videos, it looks as though several words are dropped in subtitle sentences. This often appears in Japanese variety shows.
\end{itemize}

The participants of {\it Alignment task} of the MGB challenge\cite{bellSystemForAutomaticAlignment2015, lanchantinSelectionOfMultiGenre2016, manoharJHUKaldiSystem2017} discarded the temporal information of subtitles and used the so-called {\it lightly-supervised} approach.
After applying speech recognition to audio using a biased language model (LM), which is a language model adapted to subtitle texts, output transcripts and subtitle texts are forced-aligned to select matching segments.
However, when we simply employed the {\it lightly-supervised} method as done in previous studies\cite{manoharJHUKaldiSystem2017,hernandezTEDLIUMTwiceMuch2018} in our task, the average extraction rate was not as high as we desired.
The average word extraction rate of all programs was 68\%, which further decreased to 63\% when overly short segments of less than a second were removed. 
This was only 45\% in variety shows after the removal.
We presumed that this was mainly due to the several aforementioned peculiar characteristics of Japanese TV programs when compared to audiobooks or TV programs used in the previous studies.
%

The authors in \cite{bangAutomaticConstructionLargeScale2020} tried to utilize and refine the subtitle timestamps by concatenating the adjacent subtitles, expanding the temporal boundaries forward and backward for a few seconds, and applying the forced alignment.
However, this approach cannot deal with realtime subtitles, in which the deviation could be far more than a few seconds.

Our approach is generally based on the lightly-supervised decoding approach.
We specifically adopted the scripts provided in the Kaldi ASR toolkit\cite{poveyKaldiSpeechRecognition2011}\footnote{\url{https://github.com/kaldi-asr/kaldi/blob/master/egs/wsj/s5/steps/cleanup/{segment_long_utterances_nnet3.sh,clean_and_segment_data_nnet3.sh}}}, which is based on the method described in \cite{manoharJHUKaldiSystem2017}.
We proposed the iterative alignment of the remaining unmatched segments to extract the segments missed in the previous alignment attempts to overcome the low extraction rate issue.
%
Our approach consisted of the following five steps as shown in \figref{fig:diagram}.
\begin{description}
\item[Step 1] For each recorded program, the entire audio and the corresponding concatenated subtitles were aligned using the result of the ASR with a biased LM trained with subtitle texts. We first decoded the audio to obtain a text transcript with timestamps using a speech recognizer.
We then aligned them with the original subtitle texts using the Smith--Waterman alignment. \texttt{segment_long_utterances_nnet3.sh} script of Kaldi was used here.

\item[Step 2] The ASR with another biased LM trained only with subtitle texts corresponding to the segment was applied for each segment extracted in the previous step.
The invalid portions of the transcript of each segment were then removed based on the recognition results.
\texttt{clean_and_segment_data_nnet3.sh} script of Kaldi was used here.
We then removed the segments with transcripts having less than 10 words.
The outputs of this step contained decoded transcripts, not subtitle texts.

\item[Step 3] We obtained segments with timestamps and corresponding subtitle texts by aligning the decoded transcripts of the extracted segments with the original subtitle.
By this step, we can also extract segments that were not aligned.
We used the \texttt{SequenceMatcher} module in Python's \texttt{difflib} package for the text alignment.
We executed the alignment process in both forward and backward directions and treated the tokens aligned with exactly the same token in both directions as validly aligned tokens.
\item[Step 4] For the segments that are not aligned in the previous step, we executed the same ASR-based data cleansing as in Step 2.
We did not remove segments based on the number of words in this step.
\item[Step 5] We merged the segments extracted by steps 3 and 4.
We then passed the resultant dataset to Step 3 as ``Aligned data'' in \figref{fig:diagram}, and repeated steps 3--5.

\end{description}

We expected that the repetition of steps 3--5 would allow us to achieve more data, especially those relatively difficult to align correctly in the first two steps, and contribute to the acoustic and linguistic variety of the corpus.
In our current implementation we set the maximum number of repetitions to 2.
After the repetitions, we passed the ``Merged data'' from the last Step 5 to Step 3 and obtained the final validated dataset.
We removed the segments shorter than a second to ensure that all segments had a certain audio length for the acoustic model training.

For the abovementioned decoding steps, we used Kaldi's nnet3 acoustic model (AM) with a TDNN-chain architecture trained on the CSJ corpus, containing approximately 520 hours of academic speech dialogs.
The model was built using Kaldi's official CSJ recipe \footnote{\url{https://github.com/kaldi-asr/kaldi/tree/master/egs/csj/s5}}.

\section{Details of the corpus}
\label{sec:corpus_details}

\subsection{Statistics}
\label{sec:corpus_statistics}

\tabref{tab:genre_statistics} shows the statistics of the raw recordings and the extracted corpus.
For simplicity, we only used the first genre tag obtained from each recording, even though some programs belonged to multiple genres.
The word coverage rate was calculated as (total \#words in the extracted segments) / (total \#words in the original subtitles).
We did not calculate the coverage rates in terms of the audio lengths because they were influenced by factors, such as the total lengths of commercials in a program.

The extraction process worked properly because the coverage rates were higher than 50\% in all genres.
More precisely, the word extraction rate largely varied between the genres from the lowest 53.2 \% in variety shows to the highest 89.9 \% in news reports.
This variation between genres seems to be caused by the difference of quality of audio signals (i.e the level of background noises, spontaneousness, etc.)
A noticeable deviation in the audio lengths between the genres in the final corpus was found due to such differences and the inequal total lengths of the raw recordings of each genre.
Nevertheless, the corpus contained speeches of a large variety of speakers and acoustic environments, which is beneficial to building an ASR system.

\subsection{Released Data}
\label{sec:corpus_data}

All audio samples were segmented based on the boundaries of the original subtitles, with sampling rates of \SI{16}{kHz}.
The order of all segments was randomly shuffled, regardless of their original TV programs
to prevent the corpus from being listened to as a TV program.
The transcripts were in the form of a sequence of tokenized words.
Each word contained a simple morpheme tag such as ``noun'' or ``verb,'' obtained through preprocessing the original subtitles with MeCab.

We split the whole dataset into training and development sets (\tabref{tab:corpus_splits}).
We randomly selected 1,000 speech segments from each genre listed in \tabref{tab:genre_statistics} and defined the resultant dataset of 13.7 hours of speech as the development data.
The training data comprised the remaining dataset of 2036.2 hours.
We did not provide a test set in this corpus because it was nearly impossible not to make speakers in the test set appear in the training or development set considering that many celebrities appear in several programs, and we cannot obtain the exact information of the casts of each program.
Instead of the test set of LaboroTVSpeech, we provided TEDxJP-10K, which is a different dataset built using Japanese TEDx videos.
The details of the dataset is further described in \ref{sssec:ted10K}.

\section{AUTOMATIC SPEECH RECOGNITION EXPERIMENTS}
\label{sec:asr}

\subsection{Experimental Setup}
\label{ssec:exp_setup}

We evaluated our corpus through two speech recognition tasks using two evaluation datasets: the TEDxJP-10K and CSJ evaluation datasets, (i.e., eval 1--3).
TEDxJP-10k is a dataset we built to evaluate ASR corpora, which is further described in \ref{sssec:ted10K}.
We report the character error rate (CER) in each task using all words in the reference and hypothesized transcripts, including fillers and other words caused by disfluencies.

\subsubsection{The TEDxJP-10K dataset}
\label{sssec:ted10K}

The TEDxJP-10k dataset was built to compare the performances of the Japanese ASR systems under a more realistic situation with diversed recording conditions, dialects and background noises, etc.
In many machine learning datasets, the test sets are developed as subsets of the entire data, which results in very similar characteristics between the train and test sets.
This is not a problem concerning the comparison of the performances of various models.
However, when the evaluation purpose is to compare the dataset performances, this is not ideal because the performance of a model trained on data similar to test data will be naturally higher, hindering a fair comparison.
Thus, we decided to build an independent test dataset to fairly evaluate the performance of our corpus. 

We built the dataset by first collecting the audio and the subtitles from the videos listed in the ``TEDx talks in Japanese'' playlist\footnote{\url{https://www.youtube.com/playlist?list=PLsRNoUx8w3rOHjXIU5EE4KOiIagv9yQaG}}.
For each video with manual (i.e., not auto-generated) subtitles, we segmented the audio at the subtitle boundaries.
We randomly selected 10,000 segments from all the audio segments of all talks.
An annotator manually checked and modified each of them.
During this step, the timestamps were modified if they were inaccurate. Moreover, the transcripts were re-written to exactly match the spoken words, including fillers, if necessary.
The dataset contained 10,000 audio segments spoken by 273 unique speakers of different genders, ages, and backgrounds.
The total audio length was 8.8 hours.
We released the list of video URLs and the annotator's modification; hence, exactly the same data can be reconstructed\footnote{\url{https://github.com/laboroai/TEDxJP-10K}}.

\subsubsection{Acoustic and Language Models}
\label{sssec:exp_models}

We compared three different AMs, namely CSJ, TV, and CSJ+TV.
The CSJ model was the same as that used in \secref{ssec:align}.
The TV model was trained on our new corpus.
The CSJ+TV model was trained on both CSJ and TV corpora by simply combining them.
All AMs were trained with the same procedures as Kaldi's official CSJ recipe.
We did not conduct any parameter tuning specific to our corpus to compare only the data difference.
For the TV data, we did not have speaker labels for each audio segment.
Therefore we treated each speech segment as an utterance spoken by a unique speaker, including when computing the cepstral mean and variance normalization (CMVN), training GMMs with speaker adaptive training (SAT), and training $i$-vector extractors.

We used 3-gram LMs trained with modified Kneser--Ney discounting using SRILM\cite{stolckeSRILMEXTENSIBLELANGUAGE2002} in all experiments following the default CSJ recipe.
The LM built in the CSJ recipe (CSJ-LM) was used for all the AMs to decode the CSJ evaluation dataset.
We tested the LMs trained using two different text corpora in addition to the CSJ-LM to decode the TEDxJP-10K dataset.
The TV-LM was built using the transcripts of the training set of LaboroTVSpeech.
The vocabulary size of the TV-LM was approximately 170K.
We also built LMs using the OSCAR\cite{suarezAsynchronousPipelineProcessing2019} dataset containing 100 GB of web-crawled Japanese texts.
We cleansed the data by removing special symbols such as \textit{emoji}s, and tokenized the whole data in the same procedures as described in \secref{sec:collection}.
In building OSCAR-LM, we selected 200K vocabulary based on the frequency counts of words. Subsequently, 2 and 3-gram probabilities were pruned with a threshold of $10^{-8}$.
We also examined the interpolating TV-LM and OSCAR-LM (TV+OSCAR).
Their interpolation weights were estimated with an expectation--maximization (EM) algorithm using the transcripts of the development data.

\subsection{Experimental Results}
\label{ssec:exp_results}


\begin{table}
  \center
  \caption{ASR decoding results (CER\%) on the TEDxJP-10K dataset.}
  \label{tab:asr_results_ted}
  \begin{tabular}{l||ccc}
    \hline
             & \multicolumn{3}{c}{Acoustic model                }                                                       \\
    LM       & \multicolumn{1}{c}{CSJ}                            & \multicolumn{1}{c}{TV} & \multicolumn{1}{c}{CSJ+TV} \\
    \hline
    CSJ      & 23.41                                              & 21.83                      & 20.61                          \\
    TV       & 22.14                                                  & 18.98                      & 18.22                          \\
    OSCAR    & 23.52                                              & 19.28                      & 18.71                          \\
    TV+OSCAR & 21.89                                                  & 18.29                      & 17.92                          \\
    \hline
  \end{tabular}
  \vspace{-.5\baselineskip}
\end{table}

\tabref{tab:asr_results_ted} presents the ASR performance in the CER for the TEDxJP-10K dataset.
Comparing CSJ and TV, TV consistently showed better performances.
When TV+OSCAR-LM was used, TV-AM improved CER by 3.6\% points from CSJ-AM, clearly showing that the AM built using our corpus is more capable of recognizing speech under realistic conditions.
Moreover, combining the training data of the CSJ and our TV corpus showed a noticeable decrease in the CER.
The improvement from CSJ-AM was 4.6\% points when TV+OSCAR-LM was used.

\begin{table}
  \center
  \caption{ASR decoding results (CER\%) on the CSJ eval dataset using CSJ-LM.}
  \label{tab:asr_results_csjeval}
  \begin{tabular}{l||ccc}
    \hline
            & \multicolumn{3}{c}{Acoustic model                }                                                       \\
    Dataset & \multicolumn{1}{c}{CSJ}                            & \multicolumn{1}{c}{TV} & \multicolumn{1}{c}{CSJ+TV} \\
    \hline
    eval1   & 8.00                                               & 14.46                       & \textbf{7.83}                          \\
    eval2   & \textbf{6.44}                                      & 11.30                      & 6.62                          \\
    eval3   & \textbf{5.94}                                      & 10.94                      & 6.22                          \\
    \hline
  \end{tabular}
  \vspace{-.5\baselineskip}
\end{table}

\tabref{tab:asr_results_csjeval} shows the ASR performance in the CER for the CSJ eval dataset.
The CSJ-AM showed a better performance than TV-AM because the acoustic domain of the CSJ eval dataset matched that of the CSJ training data, as discussed in \secref{sssec:ted10K}.
However, by combining the two corpora (CSJ+TV), the results significantly approached those of the CSJ-AM.
In eval1, CSJ+TV provided a slight improvement of 0.17\% points.

\tabref{tab:asr_results_ted} and \tabref{tab:asr_results_csjeval} show that CSJ+TV-AM exhibited promising results in both ASR tasks.
These results demonstrate that our new corpus is useful for training an ASR system by itself, or even in combination with other corpora.

\section{Conclusions}
\label{sec:conclusions}

This study introduced LaboroTVSpeech, a new Japanese ASR corpus containing over 2,000 hours of various speeches.
The corpus was built using Japanese broadcast TV recordings and their subtitles, which are not always verbatim.
The corpus was made available for academic usage.
We described the statistical details of the corpus and evaluated its practicality through two ASR tasks.
The experiment results showed that our new corpus is especially useful for building ASR systems that can recognize speech with various acoustic characteristics.

One key advantage of our corpus is that its size increases as time goes on because we use broadcast TV programs as data sources.
Therefore we plan to regulary update the corpus with a larger amount of data.
Moreover, the genre information on TV recordings allows us to make subsets of the whole dataset containing speech segments of a specific genre.
Such sub-datasets would be useful particularly in building an ASR system for a specific domain.



\bibliographystyle{IEEEbib}
\bibliography{strings,refs}

\end{document}